\documentstyle[prl,aps]{revtex} 
\tighten
\input psfig.tex
\begin{document}

\title{Scaling treatment of the random-field Ising model}
\author{R. B. Stinchcombe$^a$,
E. D. Moore$^a$\footnote{Electronic address: dmoore@thphys.ox.ac.uk}
and S. L. A. de Queiroz\footnote{Electronic address: sldq@portela.if.uff.br}
 }
\address{
$^a$ Department of Physics, Theoretical Physics, University of Oxford,\\
 1 Keble Road, Oxford OX1 3NP, United Kingdom\\
$^b$Instituto de F\'\i sica, Universidade Federal Fluminense,\\ Outeiro de
S\~ao Jo\~ao Batista s/n, 24210-130 Niter\'oi RJ, Brazil}
\date{\today}
\maketitle
\begin{abstract}
Analytic phenomenological scaling is carried out for the random field Ising
model in general dimensions $d$ using a bar geometry. Domain wall
configurations
and their decorated profiles and associated wandering and other
exponents $(\zeta,\gamma,\delta,\mu)$ are
obtained by free energy minimization. Scaling between different bar widths
provides the renormalization group (RG) transformation. Its consequences are
$(i)$ criticality at $h=T=0$ in $d \leq 2$ with correlation length $\xi(h,T)$
diverging like $\xi(h,0) \propto h^{-2/(2-d)}$ for $d<2$ and $\xi(h,0) \propto
\exp[1/(c_1\gamma h^{\gamma})]$ for $d=2$, where $c_1$ is a decoration
constant; $(ii)$ criticality in $d = 2+\epsilon$ dimensions at $T=0$,
$h^{\ast}=
(\epsilon/2c_1)^{1/\gamma}$, where $\xi \propto
[(s-s^{\ast})/s]^{-2\epsilon/\gamma}$, $s \equiv h^{\gamma}$. Finite
temperature
generalizations are outlined. Numerical transfer matrix calculations and
results from a ground state algorithm adapted for strips in $d=2$ confirm the
ingredients which provide the
RG description.
\end{abstract}

\pacs{PACS numbers:  75.10.Nr, 05.50+q}
\twocolumn
The random field Ising model (RFIM), which is closely related to dilute
antiferromagnetic Ising systems in a uniform field\cite{fis79}, has provided
a sequence of challenges to theoreticians as well as
experimentalists\cite{yos82,bir85,bel83,aha76}. The eventual
resolution~\cite{imb84} of the puzzle concerning its lower critical dimension
confirmed the validity of domain wall pictures\cite{imr75}. Attempts to
elucidate the critical behaviour at the lower critical dimension, $d=2$,
by normally very powerful numerical techniques\cite{pyt85,fer85,gla86}
have encountered difficulties related to the apparently anomalously severe
divergence of the correlation length there. The phase diagram boundaries are
not known for general dimensions, and the critical behaviour in three
dimensions\cite{you93,gof93} is presently not understood.

Yet methods for describing individual domain walls have been
provided\cite{der83,nat85,vil82,bin83}, and their exploitation in standard
geometries\cite{vil82,bin83} has been very effective, within the limits of a
scheme without renormalization.

The aim of the present work is to use domain wall pictures in a bar geometry
to build up, through finite size scaling\cite{bar83,nig90}, a renormalization
group
(RG) description from which the critical properties can be obtained. For strip
geometries in $d=2$, we also present preliminary free energy results from a
numerical transfer matrix procedure~\cite{ourtm} from which quantities such as
the free energy and correlation length can be estimated. Though presently short
of providing a full self-contained numerical phenomenological RG description,
these together with and a direct numerical solution for the ground
state~\cite{ogi86a}
allow a check of the analytical domain wall arguments from which
the renormalization scheme is built up.

The method exploits RG transformations obtained by finite size scaling in the
bar geometry shown in Fig.~\ \ref{fig:bar}.

We begin with the case of zero temperature $T$, which most easily illustrates
the procedure and already includes several important results and
considerations.
Here the basic quantity is the energy $\cal U$ per site. Throughout the letter
the no-wall term is subtracted, and  hypercubic or continuum cases are
considered.
For flat domain walls with average spacing $\xi_L$,\, $\cal U$ becomes ${\cal
U}^F=WL^{1-d}/\xi_L$ where
\begin{equation}
W = J(2L^{d-1}-4c_{0}h(\xi_{L}L^{d-1})^{1/2})\ \ .
\label{eq:1}
\end{equation}
\noindent This is made up of a surface term and one proportional to
(volume)$^{1/2}$. The variable $h$ is the rms random field in units of the
exchange interaction $J$, $d$ is the dimensionality and $c_0$ is a constant of
order 1.
Minimizing ${\cal U}^F$ with respect to $\xi_L$ gives the equilibrium values
$\xi_{L}^{F}$, $U^F$:
\begin{equation}
\xi_{L}^{F} = L^{d-1}(c_{0}h)^{-2}\ \ , \ \ U^F = -2(c_{0}h)^{2}/L^{d-1}\ \ .
\label{eq:2}
\end{equation}
As discussed later, these primitive results are asymptotically correct
for small $h$, where $\xi_L \gg L$, and are confirmed by comparison with the
numerical transfer matrix (TM) results in the appropriate limit.

The RG transformation of parameters (here $h \rightarrow h^{\prime} =
R_{b}(h)$)
under rescaling of $L$ by $b$ is obtained from the phenomenological scaling
relation\cite{bar83,nig90}
\begin{equation}
\xi_{L}(h^{\prime})/L = \xi_{bL}(h)/(bL)\ \ .
\label{eq:3}
\end{equation}
\noindent Using $\xi_{L}^{F}$, as in Eq. \ref{eq:2}, gives
\begin{equation}
h^{\prime} = R_{b}^{F}(h) \equiv b^{(2-d)/2} h \ \ .
\label{eq:4}
\end{equation}
\noindent This result has been previously obtained in the field-theoretic
scheme of Grinstein and Ma\cite{gri82}; see also Ref. \onlinecite{bra85}.
Eq. \ref{eq:4} has an unstable fixed point at $h=0$ for $d<2$. Here the
eigenvalue $b^{(2-d)/2}$ implies that the bulk correlation length diverges for
small $h$ at $T=0$ in $d<2$ like
\begin{equation}
\xi \propto h^{-2/(2-d)}\ \ .
\label{eq:5}
\end{equation}

The transformation (\ref{eq:4}) is marginal for $d=2$. So, subdominant terms
are
required, even at small $h$, $T=0$, for $d=2$, to deal with the marginality.
These come from ``decoration'' perturbations of the flat
walls\cite{nat85,vil82,bin83}. These can be built up from smooth shape
modifications of base length $a$ and height $b(a)$, superimposed for all scales
$a$ between $L$ and a cutoff value determined by the lattice constant (=1),
see Fig.~\ \ref{fig:dec} .
The minimum energy $\varepsilon (a)$ and minimizing height $b$ for a smooth
decoration of base scale $a$ are of the form
\begin{equation}
\varepsilon (a) \propto - h^{\gamma} a^{\delta}\ \ , \ \ b(a) \propto
h^{\mu} a^{\zeta}\ \ ,
\label{eq:6}
\end{equation}
\noindent where $\zeta$ is a wandering exponent\cite{nat85,gri82,halpin,mez90},
and the exponent $\gamma$ will
be particularly significant below. The minimization gives (general $d$):
\begin{equation}
(\gamma ,\delta ,\mu ,\zeta ) = ({4 \over 3}, {d+1 \over 3}, {2 \over 3}, {5-d
\over 3} ) \ \ {\rm or} \ \
(2, 1, 2, 3-d )\ \ ,
\label{eq:7}
\end{equation}
\noindent where the first result applies for the continuum case if $b \ll a$,
and the second one applies to the continuum if $b \ge a$, and to the lattice
case. So, in the continuum case, if $a$ is large the first result applies for
$d>2$ (largest $\delta$, $i.e.$ lowest energy) and the second for $d<2$.

Since in Eq.  \ref{eq:6} $\varepsilon (a)$ is negative, decorations on
decorations occur down to a smallest base length $a_m$ such that $\min
(a_m~,b(a_m)) \approx 1$.  The resulting contribution of all
decorations to the energy
of a single wall of base length $L$ is, for $d=2$,
\begin{equation}
\Sigma = -Jc_1 h^{\gamma} L \ln (L h^{\mu} c_2 )\ \ ,
\label{eq:8}
\end{equation}
\noindent where $c_1$ and $c_2$ are constants of order 1. The effect of
decorations  is therefore to replace the wall energy $W$ by $W + \Sigma$.
Minimizing the energy per site with respect to wall separation then gives
equilibrium values $U^D$, $\xi_{L}^{D}$ for the decorated wall generalization,
and hence (via $\xi_{L}^{D}/L$) the generalized phenomenological scaling
equation. For $d$ near 2 (the marginal case), this is:
\begin{equation}
h^{\prime} = h b^{(2-d)/2} [1+c_1 h^{\gamma} \ln b]\ \ .
\label{eq:9}
\end{equation}
\noindent From Eq. \ref{eq:7}, $\gamma = 4/3$, 2 for continuum and lattice
cases
respectively. The transformation given by Eq. \ref{eq:9} has the proper
semigroup character,
in its range of validity ($h$ small, $d$ near 2, $T=0$). As will be seen, this
range includes the critical effects for  $d$ at or just above 2.

For $d=2$, Eq. \ref{eq:9} shows that the $h=0$ fixed point is marginally
unstable, as expected at the lower critical dimension. Analyzing the equation
by standard RG procedures gives the two-dimensional bulk correlation length
diverging for small $h$ and $T=0$ as
\begin{equation}
\xi = A \exp{\Bigl({1 \over c_1 \gamma h^{\gamma}}\Bigr)}\ \ .
\label{eq:10}
\end{equation}
In $2 + \epsilon$ dimensions, $\epsilon$ small and positive, the RG
transformation Eq. \ref{eq:9} yields an unstable zero temperature fixed point
at small non-zero field $h^{\ast}$, as well as the associated divergence of
$\xi$ :
\begin{equation}
h^{\ast} = (\epsilon /2 c_1)^{1/\gamma}\ \ ,\ \  \xi \propto
[(s-s^{\ast})/s]^{-2/\epsilon\gamma},\ s \equiv h^{\gamma}\ \ .
\label{eq:11}
\end{equation}
This divergence crosses over to the behaviour (\ref{eq:10}) as $\epsilon
\rightarrow 0$. In Eqs. (\ref{eq:10}) and (\ref{eq:11}) the exponent $\gamma$
differs between lattice and continuum systems. This feature has previously been
noted in the interface context\cite{nat85}.

For larger fields, the flat wall description requires more drastic
generalizations. Increasing $h$ at fixed $L$ would, according to Eq.
\ref{eq:2},
eventually give $\xi_L < L$. So, by then the effects of domain boundaries in
all
directions should have been allowed for. This generalization\cite{moore} yields
a complicated form which confirms Eq. \ref{eq:1} and its consequences in the
low
field regime where $\xi_L \gg L$, but which for $\xi_L \ll L$ gives
\begin{equation}
{\cal U} = J[c_1 \xi_{L}^{-1} - c_2 h \xi_{L}^{d(\phi -1)}]\ \ ,
\label{eq:12}
\end{equation}
\noindent where $c_1$, $c_2$ are geometric constants and $\phi$ crosses over
between 1/2 and 1 as $\xi_L$ approaches order 1. For $d=2$, $T=0$ the result of
allowing for this, and the decoration effects, is as follows. As $h$ increases
the low field result for $U$ arising from Eq. \ref{eq:2} (namely
$-2(c_{0}h)^{2}/L$) picks up a factor $\Lambda (h,L) \equiv [1+c_1~
h^{\gamma}~\ln(Lh^{\mu}c_2)]$, then goes through a complicated intermediate
regime, and finally for $h$ larger than about 1 becomes linear in $h$ and
independent of $L$.

A check of the basic (flat wall) ingredients in the domain scaling description
is provided by comparing these results for $U$ with those provided by a
numerical transfer matrix calculation for the $d=2$ RFIM\cite{ourtm}. Here the
configurationally
averaged free energy is enumerated exactly for very long ($\sim  10^5$ lattice
parameters) strips of finite width $L$. Fig. \ref{fig:num} gives the
field dependence of the negative of the free energy per spin $f$, for
relatively small widths $L=2, \dots 7$, at temperature $T/J=0.1$ which is
sufficiently
low that $f$ is essentially $U$.  The main characteristics of the domain wall
results for $-U$
($h^{2}/L$ crossing over to $h/L^0$), may be seen to be present. Indeed the
analytical results, taking $c_0 = 0.4614$ in Eq. \ref{eq:2}, fit the numerical
data for
$L = 4 - 9$ within the very small error estimated from the data fluctuations
in the low field regime $h \le 0.5$ for $T/J=0.1$, and also for higher
temperatures using extensions to the domain wall theory outlined immediately
below. However, the fit is not sufficiently sensitive to the decoration terms
to confirm the theoretical value $\gamma =2$. Other comparisons discussed later
show more direct effects of the decoration terms.

The finite temperature generalization involves entropic terms. At the most
primitive level (flat walls only) the entropy per site is:
\begin{equation}
S^F = -k_{B}L^{(1-d)}[x\ln x +(1-x)\ln (1-x)],\ x \equiv 1/\xi_L,
\label{eq:13}
\end{equation}
\noindent coming from the number of ways of selecting ${\cal L}x$ out of
${\cal L}$ sites for the placement of the walls. Adding $-TS^F$ to the flat
wall internal energy ${\cal U}^F$ and minimizing gives the equilibrium flat
wall
separation $\xi_{L}^{F}$ and the free energy $f^F$ per site. $\xi_{L}^{F}$
satisfies
\begin{equation}
c_{0} h(\xi_{L}^{F}/L^{d-1})^{1/2} + (L^{1-d}/2K)\ln(\xi_{L}^{F}-1) =1\ \ ,\
\label{eq:14}
\end{equation}
where $K \equiv J/k_{B}T$.

This result is valid if $\xi_L \gg L$, which is the case if both
$h L^{(2-d)/2}$ and $(1/K)L^{1-d}\ln L$ are small, sufficient to encompass Eqs.
 \ref{eq:2}, \ref{eq:4} and \ref{eq:5} and their low-temperature
generalizations. Otherwise domain walls in all directions are needed, and they
increase the total entropy, asymptotically to
$c_{3}\xi_{L}^{-d}\ln(c_4\xi_{L})$
where $c_{3,4}$ are constants. And in $d=2$ decoration terms are required to
resolve the marginality at low $T$. They affect both the energy as discussed
above (taking $K$ in Eq. \ref{eq:14} to $K\Lambda(h,L)$), and the entropy.
Their
entropy contribution for each vertical wall of base length $L$ in $d=2$ is of
the form $c_{5}Lh^{\mu}\ln(c_{6}L)$. The resulting temperature-dependent
generalizations of Eqs. \ref{eq:4}, \ref{eq:5}, \ref{eq:9}, \ref{eq:10} and
\ref{eq:11} will be given elsewhere\cite{moore}, together with phase boundaries
and comparisons between analytic and numerical transfer matrix and Monte
Carlo~\cite{mymc} results for
finite temperature free energies.

However, we briefly present here evidence for the marginality--breaking domain
wall roughening given in equation~(\ref{eq:6}) and used to derive key results
such as~(\ref{eq:8}),~(\ref{eq:9}),~(\ref{eq:10}) and~(\ref{eq:11}).
Specifically, we demonstrate that, for a gaussian distribution of random
fields, the exponent $\mu$ approaches $2$ as $L$ grows large.  The max--flow
algorithm of Ogielski~\cite{ogi86a}  was implemented on a strip geometry of
dimensions $L \times 1000$ and the ground state was generated for $100$
independent random field configurations of fixed standard deviation $h$.  For
each spin configuration, the rms width of the domain walls was measured, and
this statistic was averaged
over all the domain walls of the $100$ field configurations.  We expect
$b_{rms}(L)$ to scale like $b(a)$ in equation~(\ref{eq:6}).
Figure~\ref{fig:muplot} justifies this prediction for large $L$.

Detailed results, discussions of the operational procedures will be given
elsewhere\cite{ourtm,moore,mymc}.  The results are here used merely to support
the domain scaling approach.

In conclusion, an analytic phenomenological scaling approach has been
constructed using domain considerations.  The resulting RG
transformations have been used to find critical properties, including
the correlation length behaviour at and near the marginal (lower
critical) dimension $d=2$.  Numerical transfer matrix free energy comparisons
have confirmed, in $d=2$, the basic
domain picture used in the scaling. And numerical ground state wall roughening
investigations have quantitatively tested a particular theoretical prediction
(the exponent $\mu$) for the decoration ingredient, which is so crucial
for the criticality
in 2 and $2 + \epsilon$
dimensions (see Eqs. \ref{eq:10}, \ref{eq:11}).

Generalizations of the domain scaling approach to related vector model systems
and to dynamics are being pursued starting from existing domain wall pictures
(respectively Refs. \onlinecite{imr75} and \onlinecite{hus87a}).

\acknowledgements

We would like to thank M.A. Moore and J.L. Cardy for relevant comments.
SLAdQ thanks the Department of Theoretical Physics
at Oxford, where this  work was initiated, for the hospitality, and
the cooperation agreement between Academia Brasileira de Ci\^encias and
the Royal Society for funding his visit. SLAdQ also thanks Departamento de
F\'\i sica, PUC/RJ for use of their computational facilities.
Research of SLAdQ
is partially supported by the Brazilian agencies Minist\'erio da Ci\^encia
e Tecnologia, Conselho Nacional
de Desenvolvimento Cient\'\i fico e Tecnol\'ogico and Coordena\c c\~ao de
Aperfei\c coamento de Pessoal de Ensino Superior.

EDM thanks the Rhodes Trust and the Natural Sciences and Engineering Council of
Canada for funding this research.  He would also like to thank Michael Swift
for supplying the code for the max--flow algorithm.


\newpage

\begin{figure}
\caption{
Domain walls of separation $\xi_L$, in a $d$-dimensional bar of transverse
scale $L$.}
\label{fig:bar}
\end{figure}

\begin{figure}
\caption{
Domain wall decorations ($d=2$ for simplicity) of base length $a$, height
$b(a)$: $(i)$ for continuum; $(ii)$ for lattice; $(iii)$ superimposed at
successively smaller scale.}
\label{fig:dec}
\end{figure}

\begin{figure}
\caption{
Free energy for the $d=2$ RFIM, from numerical transfer matrix calculations for
long strips of widths $L=2,3, \dots 7$, as a function of $h/J$ at $T/J=0.1$.
The behaviour crosses over from $h^{2}/L$ to $h/L^0$, as predicted by domain
wall arguments and used in scaling.}
\label{fig:num}
\end{figure}

\begin{figure}
\caption{
The domain wall roughening exponent $\mu$.  This has been determined by
least-squares fits to semi--log plots of $b_{rms}(L)$ vs $h$.}
\label{fig:muplot}
\end{figure}

\end{document}